\documentclass[prl,showpacs,twocolumn,aps,superscriptaddress,nofootinbib]{revtex4-2}
\usepackage{amssymb,amsmath,bm}
\usepackage{graphicx}
\usepackage{mathpazo}
\usepackage[colorlinks=true, pdfstartview=FitV, linkcolor=blue, citecolor=blue, urlcolor=blue]{hyperref}

\newcommand{\bb}{{\boldsymbol b}}

\newcommand{\bx}{{\boldsymbol x}}

\newcommand{\bk}{{\boldsymbol k}}
\newcommand{\bv}{{\boldsymbol v}}

\newcommand{\bB}{{\boldsymbol B}}
\newcommand{\bE}{{\boldsymbol E}}
\newcommand{\be}{{\boldsymbol e}}

\newcommand{\bj}{{\boldsymbol j}}
\newcommand{\bnabla}{{\boldsymbol \nabla}}

\newcommand{\bo}{{\boldsymbol \omega}}

\renewcommand\a{\alpha}

\renewcommand\d{\delta}
\renewcommand\k{\kappa}

\renewcommand\r{\rho}
\renewcommand\t{\tau}

\renewcommand\c{\chi}

\renewcommand\o{\omega}

\newcommand\m{\mu}

\newcommand\x{\xi}
\newcommand\p{\pi}

\newcommand\s{\sigma}

\newcommand\w{\eta}
\newcommand\ve{\varepsilon}

\newcommand\G{\Gamma}

\newcommand\ra{\rightarrow}

\newcommand\pt{\partial}

\newcommand\lb{\left(}
\newcommand\rb{\right)}

\newcommand{\lan}{\langle}
\newcommand{\ran}{\rangle}

\newcommand{\eq}[1]{Eq.~(\ref{#1})}
\newcommand{\eqs}[2]{Eqs.~(\ref{#1})-(\ref{#2})}
\begin{document}

\title{Chiral magnetovortical instability}
\author{Shuai Wang}
\affiliation{Physics Department and Center for Particle Physics and Field Theory, Fudan University, Shanghai 200438, China}
\author{Xu-Guang Huang}
\email{huangxuguang@fudan.edu.cn}
\affiliation{Physics Department and Center for Particle Physics and Field Theory, Fudan University, Shanghai 200438, China}
\affiliation{Key Laboratory of Nuclear Physics and Ion-beam Application (MOE), Fudan University, Shanghai 200433, China}
\affiliation{Shanghai Research Center for Theoretical Nuclear Physics, National Natural Science Foundation of China and Fudan University, Shanghai 200438, China}

\begin{abstract}
We demonstrate that in a chiral plasma subject to an external magnetic field, the chiral vortical effect can induce a new type of magnetohydrodynamic instability which we refer to as the {\it chiral magnetovortical instability}. This instability arises from the mutual evolution of the magnetic and vortical fields. It can cause a rapid amplification of the magnetic fields by transferring the chirality of the constituent particles to the cross helicity of the plasma. 
\end{abstract}
\maketitle

{\bf Introduction ---} Magnetic fields and fluid vortices coexist in many physical systems, encompassing the electromagnetic (EM) plasmas in stellar and planetary objects, the quark-gluon plasma (QGP) formed in heavy-ion collisions, and the electroweak plasmas in supernovas and early Universe. Notably, the magnetic field and fluid vorticity can induce anomalous, parity-breaking, transport phenomena when the plasma is chiral, that is, when the constituent particles exhibit asymmetries between their left-handed and right-handed species. Two prominent examples are the chiral magnetic effect (CME)~\cite{Vilenkin:1980fu,Kharzeev:2007jp, Fukushima:2008xe} and the chiral vortical effect (CVE)~\cite{Vilenkin:1979ui,Erdmenger:2008rm,Banerjee:2008th,Son:2009tf}, which lead to electric currents along the magnetic and vortical fields, respectively. They both emerge from the underlying chiral anomaly, which connects fermion chirality with the topology of EM and gravitational fields. In recent years, the CME and CVE have attracted considerable attention in theoretical and experimental research across many subfields of physics, including nuclear physics, astrophysics, cosmology, and condensed matter physics; see Refs.~\cite{Huang:2015oca,Liu:2020ymh,Kharzeev:2020jxw,Chernodub:2021nff} for reviews.  

In the hydrodynamic regime (i.e., low energy and long wavelength regime), the chiral plasma can be described by the so-called anomalous hydrodynamics or chiral magnetohydrodynamics (MHD) which extends the standard MHD by incorporating the electric currents from CME and CVE. New wave modes can appear in chiral MHD, such as the chiral magnetic wave~\cite{Kharzeev:2010gd}, chiral vortical wave~\cite{Jiang:2015cva}, chiral electric wave~\cite{Huang:2013iia}, chiral Alfv\'en wave~\cite{Yamamoto:2015ria}, and chiral heat wave~\cite{Chernodub:2015gxa}. Furthermore, the CME can induce a novel magnetic-field instability (and its variants) known as chiral plasma instability or chiral dynamo instability~\cite{Joyce:1997uy,Akamatsu:2013pjd}, which activates a dynamo mechanism (i.e., the amplification of a weak seed magnetic field) similar to the $\a$-dynamo but without requiring the presence of finite mean kinetic helicity. This has profound implications for our understanding of magnetic field formation and evolution in various contexts, such as the early Universe~\cite{Joyce:1997uy,Giovannini:1997eg,Boyarsky:2011uy,Boyarsky:2012ex,Tashiro:2012mf,Dvornikov:2016jth,Gorbar:2016klv,Brandenburg:2017rcb,Schober:2017cdw}, supernovas and neutron stars~\cite{Ohnishi:2014uea,Yamamoto:2015gzz,Grabowska:2014efa,Dvornikov:2014uza,Schober:2017cdw,Sigl:2015xva,Matsumoto:2022lyb,Schober:2021yav,Brandenburg:2023aco}, QGP in heavy-ion collisions~\cite{Akamatsu:2015kau,Hirono:2015rla,Xia:2016any,Tuchin:2017vwb,Schlichting:2022fjc}, and even Weyl/Dirac semimetals~\cite{Qiu:2016hzd,Galitski:2018,Amitani:2022pmu}. Various other CME-leading instabilities were also discussed in literature~\cite{Yamamoto:2015gzz,Hattori:2017usa,Yamamoto:2021gts}.
\begin{figure}[t]
	\begin{center} 
		\includegraphics[width=0.9\hsize]{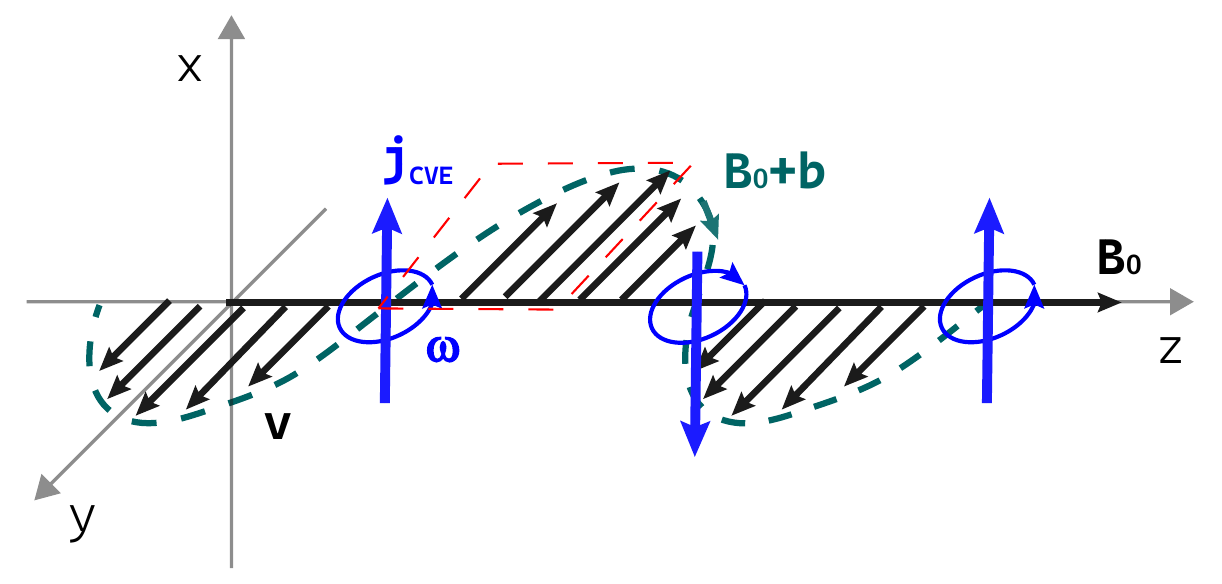}
	\end{center}
\caption{Illustration of the arising of the CMVI.
}
\label{fig:illu}
\end{figure}

Unlike the CME, the influence of CVE on the evolution of chiral plasma remains relatively unexplored. This might be because the fluid vorticity is usually considered weak comparing to the magnetic field (unless the system is in a strong kinetic-helicity dominated turbulence). However, this may not be the case when the CVE can cause or catalyze plasma instabilities. Furthermore, there are instances, such as the QGP in heavy-ion collisions, where extremely strong vorticity can appear even in laminar flow~\cite{Deng:2016gyh,Deng:2020ygd,Jiang:2016woz}. In this paper, we demonstrate that the CVE can indeed induce a new MHD instability, which we refer to as the {\it chiral magnetovortical instability} (CMVI), in a magnetized chiral plasma. 

Before going into the detailed calculation, let us provide an intuitive understanding of the CMVI. Suppose a chiral plasma is situated in a background magnetic field $\bB_0$ along the $z$ direction. Let us consider a sine-shaped perturbation of the fluid velocity $\bv$ perpendicular to $\bB_0$, as depicted in Fig.\ref{fig:illu}. In the absence of electric resistivity $\eta$, such a perturbation would cause a bending of the magnetic field line (according to Alfv\'en's frozen-in theorem), resulting in the generation of a perturbed magnetic field $\bb$ perpendicular to $\bB_0$. However, the presence of a finite $\eta$ would eventually dampen $\bb$ due to magnetic diffusion. This scenario changes when the CVE is taken into account. The perturbation in the fluid velocity induces alternating CVE currents along the $y$ direction and positioned along the $z$ axis. These electric currents generate additional magnetic fields that add to the perturbed magnetic field $\bb$. When the CVE is sufficiently strong, the perturbed field $\bb$ in regions like the one marked in the figure is amplified instead of being damped, which in turn leads to an amplification of the velocity $\bv$, leading to the emergence of an instability.

We now give an analysis using chiral MHD. We set $c=\hbar=k_B=1$ and also $\ve_0=\mu_0=1$, the charge is $q=1$.

{\bf Chiral magnetovortical instability in chiral MHD ---} We consider a chiral plasma in which the electric current $\bj$ is given by the following constitutive relation,
\begin{align}
 \label{eq:const}
\bj =n\bv+\s(\bE+\bv\times\bB)+\bj_B+\bj_\o,
\end{align}
where $\bj_B=\xi_B\bB$ and $\bj_\o=\xi_\o\bo$ ($\bo=\bnabla\times\bv$ is the vorticity) represent the CME and the CVE, respectively, with $\xi_B\propto\m_5$ and $\xi_\o\propto\m_5\m$ ($\m$ and $\m_5$ are the electric and chiral chemical potentials) the corresponding conductivities, $\s>0$ is the usual electric conductivity, and $n$ is the charge density. For the purpose of analysis, we focus on the non-relativistic limit as it offers a more transparent understanding of the CMVI, although a similar analysis can be adapted for relativistic case as well. The governing equations in the non-relativistic limit are given by the following set of chiral MHD equations (See Appendix \ref{app:derivation} for a derivation):
\begin{gather}
 \label{eq:mhd:ns}
\r\lb\pt_t +\bv\cdot\bm\nabla\rb\bv = -\bnabla P+(\bnabla\times\bB)\times\bB ,\\
 \label{eq:mhd:mag}
\pt_t\bB=  \bnabla\times(\bv\times\bB)+\eta\nabla^2 \bB +\eta\xi_B\bnabla\times\bB +\eta\xi_\o\bnabla\times\bo,
\end{gather}
accompanied by the solenoidal conditions for velocity $\bv$ (incompressibility) and magnetic field $\bB$ (Gauss law):
\begin{gather}
 \label{eq:imcom}
\bnabla\cdot\bv=\bm 0 ,\\
 \label{eq:gau}
 \bnabla\cdot\bB=\bm 0.
\end{gather}
In the above equations, $\r$ is the mass density, $P$ is the pressure, and $\eta=1/\s$ is the electric resistivity. In \eqs{eq:mhd:ns}{eq:mhd:mag}, we have neglected the viscous terms for the sake of simplicity. However, the effects of viscosity can be readily taken into account. Note that the electric field is not dynamical in MHD due to the screening effects (i.e., the timescale of MHD processes is much longer than the screening time of electric field), but determined by the constitutive relation (\ref{eq:const}).

The wave modes and possible instabilities arising from the CME have been discussed extensively. Therefore, our focus here is on the CVE. For clarity, we deactivate the CME and assume constant pressure and mass density (and thus constant $\eta, \xi_\o$) for the moment. We examine the behavior of small fluctuations around a static equilibrium state in the presence of a background magnetic field $\bB_0$, i.e., $\bv=0+\bv$ and $\bB=\bB_0+\bm b$ with $\bv$ and $\bb$ counted as of order $\d\ll 1$. 

We keep terms linear in $\d$ in \eqs{eq:mhd:ns}{eq:mhd:mag} and obtain
\begin{gather}
 \label{eq:mhd:ns2}
\r\pt_t\bv = \bB_0\cdot\bnabla{\bm b}-\bnabla(\bB_0\cdot\bb)\; ,\\
 \label{eq:mhd:mag2}
 \pt_t\bb=\bB_0\cdot\bnabla\bv+\eta\nabla^2\bb-\eta\xi_\o\nabla^2\bv.
\end{gather}
Contracting $\bB_0$ with \eq{eq:mhd:ns2} implies $\pt_t(\bB_0\cdot\bv)=0$, indicating that the longitudinal velocity fluctuation is not dynamical. Therefore, we pay our attention on the transverse velocity fluctuation by assuming $\bv\cdot\bB_0=0$. With this and the solenoidal conditions (\ref{eq:imcom}) and (\ref{eq:gau}), \eqs{eq:mhd:ns2}{eq:mhd:mag2} further imply $\pt_t(\bB_0\cdot\bb)=0$, meaning that the longitudinal magnetic field fluctuation is not dynamical. Hence, we assume $\bb\cdot\bB_0=0$ in our analysis. 

To find the eigen modes of \eqs{eq:mhd:ns2}{eq:mhd:mag2}, we substitute the plane-wave form of the fluctuations,
\begin{align}
 \label{eq:plane}
\bv={\bm f}_v e^{i(\bk\cdot\bx-\o t)},\bb={\bm f}_b e^{i(\bk\cdot\bx-\o t)},
\end{align}
where ${\bm f}_{v,b}$ are amplitude vectors, and obtain
\begin{gather}
 \label{eq:mhd:ns3}
\r\o{\bm f}_v = -(\bB_0\cdot\bk){\bm f}_b\; ,\\
 \label{eq:mhd:mag3}
 (\eta\bk^2-i\o){\bm f}_b=(\eta\xi_\o\bk^2+i\bB_0\cdot\bk){\bm f}_v.
\end{gather}
We obtain immediately the following equation for dispersion relations:
\begin{align}
 \label{eq:disrel}
\o^{2}+ i \eta \bk^{2} \o +i \eta \xi_\o \bk^{2}\frac{\bm{B}_{0}\cdot \bm{k}}{\r}    
              -\frac{(\bm{B}_{0}\cdot \bm{k})^{2} }{\r}=0,
\end{align}
whose solutions are given by
\begin{align} 
\label{dispercve} 
\o&=\o_\pm\\
&\equiv -i\frac{\eta}{2}\bk^{2} 
        \pm \sqrt{ \frac{(\bm{B}_{0}\cdot \bm{k})^{2}}{\r}- \frac{\eta^{2} \bk^{4}}{4}
  - i \eta \xi_\o  \bk^{2} \frac{\bm{B}_{0}\cdot \bm{k}}{ \r} }\\  
 &\approx \pm\frac{\bB_0\cdot\bk}{\sqrt{\r}}-i\frac{\eta}{2}\lb1\pm\frac{\xi_\o}{\sqrt{\r}}\rb\bk^2 +O(\bk^4),
\end{align}
where the last line is valid when $|\bk_z|, |\bk_\perp|\ll |\bB_0|/(\eta|\xi_\o|)$ and $|\bB_0|/(\eta\sqrt{\r})$. The first term, $\o=\pm\bB_0\cdot\bk/\sqrt{\r}$, represents the usual Alfv\'en wave modes propagating along and opposite to $\bB_0$. Without CVE, the presence of the electric resistivity always induce dissipative diffusion of the magnetic field. However, when the CVE is turned on in the parameter region $|\xi_\o|\gtrsim\sqrt{\r}$, one of the Alfv\'en wave modes becomes unstable. This is the CMVI, which amplifies the magnitudes of the magnetic-field and velocity fluctuations. 

We also note that in another parameter region, namely, when $|\bk|\gg |\bB_0|/(\eta|\xi_\o|)$ and $|\bB_0||\xi_\o|/(\eta{\r})$ (but $|\bk|\ll1/\eta$ should be always satisfied in order for the hydrodynamic analysis being applicable), we have
\begin{align} 
\label{dispercve2} 
\o\approx \pm \frac{\xi_\o}{\r}\bB_0\cdot\bk +O(\bk^{-2}).
\end{align}
The same dispersion relation was derived in Ref.~\cite{Yamamoto:2015ria} and was referred to as the the chiral Alfv\'en wave. However, In Ref.~\cite{Yamamoto:2015ria}, the magnetic field was considered non-dynamical, and as a consequence, it was found that the dispersion relation (\ref{dispercve2}) holds for $|\bk|\ra 0$. It is interesting to note that in the presence of a dynamical EM field, the situation is changed, and the $|\bk|\ra 0$ dispersion relation is actually given by the last line of \eq{dispercve}. 

Some comments are in order: (i) The CMVI appears when $|\xi_\o|>\sqrt{\r}$, regardless of whether we make the small wave number expansion, as we did in the last line in \eq{dispercve}. (ii) For $\bk\perp\bB_0$, we only have one dissipative mode $\o_-=-i\eta\bk^2$ for magnetic field diffusion (see \eq{eq:mhd:mag2}), while the velocity fluctuation does not propagate. (iii) The appearance of CMVI indicates that during the evolution of the system, the chirality of the constituent particles should decrease in order for $\xi_\o$ to decrease and eventually cease the instability. The continuity equation for helicity thus implies that the magnetic and/or flow helicities would increase, thereby triggering a dynamo action. We will analyze this possibility in the following. 

{\bf Fate of CMVI ---} The CMVI, once it takes place (i.e., when $\xi_\o>\sqrt{\r}$, assuming $\xi_\o >0$), cannot last forever. The system would evolve towards a state where $\xi_\o<\sqrt{\r}$, thus terminating the CMVI. Due to the conservation of electric charge, we expect that the decrease of $\xi_\o$ would be mainly due to the decrease of $\m_5$. To analyze how this occurs, we examine the chiral anomaly equation,
\begin{gather}
    \label{eq:ch:cont}
   \pt_t j^0_5+\bnabla\cdot\bj_5=C\bE\cdot\bB,
   \end{gather}
where $C$ is a constant representing the strength of the chiral anomaly (for the usual EM plasma, $C=1/2\p^2$), $\bj_5$ is the chiral current, and $j^0_5=n_5+\k_B\bv\cdot\bB+\k_\o\bv\cdot\bo$ with $n_5$ the chiral density of constituent particles, $\k_{B}\propto\m$ and $\k_\o\propto T^2$ the conductivities of chiral separation effect (CSE)~\cite{Son:2004tq,Metlitski:2005pr} and axial CVE, respectively. Assuming a homogeneity of the system and writing $n_5=\c_5 \m_5$, with $\c_5\propto T^2$ denoting the chiral susceptibility, we can derive the following evolution equation for $\m_5$: 
\begin{gather}
    \label{eq:ch:cont2}
   \c_5\pt_t \m_5=-\frac{C}{2}\pt_t {\cal H}_b-\k_B \pt_t {\cal H}_c -\k_\o \pt_t {\cal H}_v -\G\c_5\m_5,
\end{gather}
where ${\cal H}_b=\lan\bm A\cdot\bB\ran$, ${\cal H}_c=\lan\bv\cdot\bB\ran$, and ${\cal H}_v=\lan\bv\cdot\bo\ran$ are the average magnetic, cross, and kinetic helicities, respectively, with $\lan\cdots\ran\equiv V^{-1}\int d^3\bx(\cdots)$. Using $\bm A=\bB_0\times\bx/2+{\bm a}$ (with $\bm a$ is the fluctuating vector potential), and the conditions $\lan \bm a\ran=\bm 0=\lan\bv\ran$, one finds that ${\cal H}_b=\lan\bm a\cdot\bb\ran$ and ${\cal H}_c=\lan\bv\cdot\bb\ran$. We have also introduced the chirality relaxation rate $\G$ in order to account for the chirality-flipping process due to, e.g., massiveness of the particles~\cite{Grabowska:2014efa}. 

To proceed, we expand the fields in their Fourier modes,
\begin{align} 
    \label{fate:fourier:v} 
\bv(t,\bx) &=\int_\bk \bv(t,\bk)e^{i\bk\cdot\bx},\\
\label{fate:fourier:b} 
\bb(t,\bx) &=\int_\bk \bb(t,\bk)e^{i\bk\cdot\bx},
\end{align}
where $\int_\bk\equiv\int {d^3 \bk/(2\p)^3}$. For each Fourier mode, we further expand it in helicity basis with $\be_3(\bk)=\hat{\bk}\equiv\bk/|\bk|$ and $\be_\pm(\bk)$ as the right-hand and left-hand helicity basis vectors. They satisfy the following properties: $\hat{\bk}\times\be_\pm(\bk) =\mp i\be_\pm(\bk)$, $\hat{\bk}\cdot\be_\pm(\bk)=0$, and $\be_\pm(\bk)\cdot\be^*_\pm(\bk)=1, \be_\pm(\bk)\cdot\be^*_\mp(\bk)=0$. The solenoidal conditions for $\bv$ and $\bB$ imply that $\bv(t,\bk)=\sum_{s=\pm} v_s(t,\bk)\be_s(\bk)$ and $\bb(t,\bk)=\sum_{s=\pm} b_s(t,\bk)\be_s(\bk)$. Using this helicity expansion and focusing on the long-wavelength modes, we can rewrite \eqs{eq:mhd:ns2}{eq:mhd:mag2} as
\begin{align} 
    \label{fourier:eomz1} 
\pt_t z_{1\pm}(t,\bk) &= -i\o_+ z_{1\pm}(t,\bk),\\
\label{fourier:eomz2} 
\pt_t z_{2\pm}(t,\bk) &= -i\o_- z_{2\pm}(t,\bk).
\end{align}
Here, ${\bm z}_{1,2}=\sum_{s=\pm}z_{1,2s} \be_s(\bk)$ are the CVE-modified Elsasser fields~\cite{Elsasser:1950zz}, given by
\begin{align} 
    \label{ell:z1} 
z_{1\pm} &\approx \Big(1-\frac{i\w\xi'_\o\bk^2}{2\bB'_0\cdot\bk}\Big)v_\pm-\Big(1-\frac{i\w\bk^2}{2\bB'_0\cdot\bk}\Big)b'_\pm,\\
\label{ell:z2} 
z_{2\pm} &\approx \Big(1-\frac{i\w\xi'_\o\bk^2}{2\bB'_0\cdot\bk}\Big)v_\pm+\Big(1+\frac{i\w\bk^2}{2\bB'_0\cdot\bk}\Big)b'_\pm,
\end{align}
with the primed quantities being scaled as $\bB'_0=\bB_0/\sqrt{\r}, \bb'=\bb/\sqrt{\r}$, and  $\xi'_\o=\xi_\o/\sqrt{\r}$. Writing in helicity basis, the average kinetic energy per unit mass, magnetic energy per unit mass, and various helicities are expressed by ${\cal E}_v=\lan \bv^2 \ran/2=(1/2V)\int_\bk  (|v_+|^2+|v_-|^2)$, ${\cal E}_b=\lan \bb'^2 \ran/2=(1/2V)\int_\bk  (|b'_+|^2+|b'_-|^2)$, ${\cal H}_v=(1/V)\int_\bk |\bk| (|v_+|^2-|v_-|^2)$, ${\cal H}_b=(1/V)\int_\bk (|b_+|^2-|b_-|^2)/|\bk|$, and ${\cal H}_c=(1/V)\int_\bk  (v_+b_+^*+v_-b_-^*)$, respectively.

\begin{figure}[t]
	\begin{center} 
		\includegraphics[width=0.8\hsize]{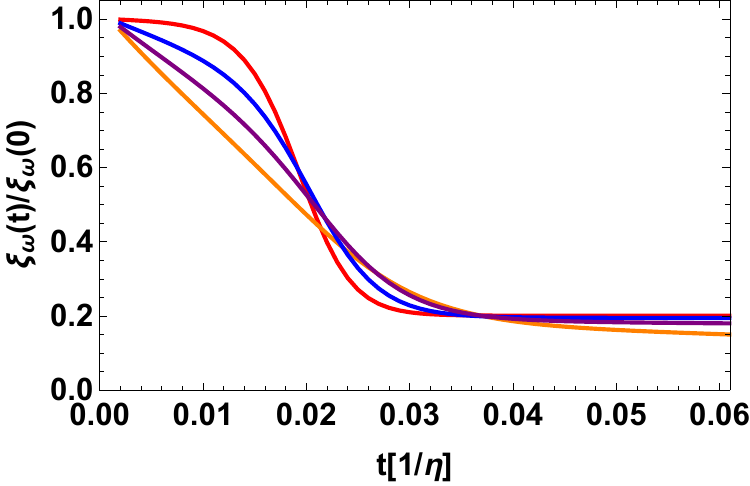}
        \includegraphics[width=0.8\hsize]{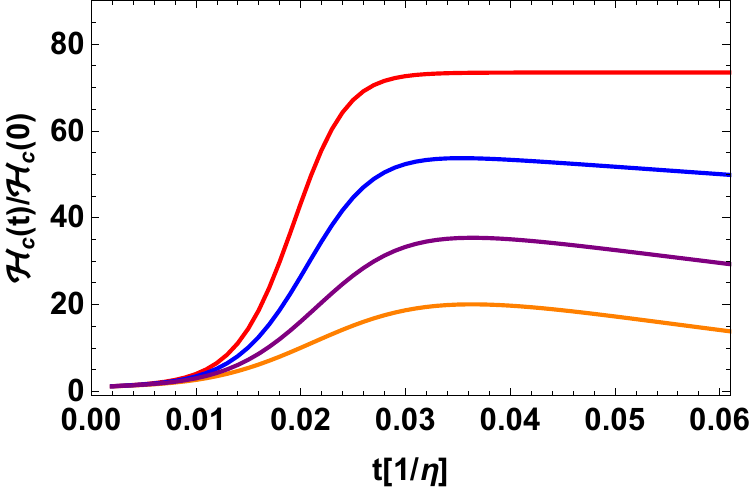}
        \includegraphics[width=0.8\hsize]{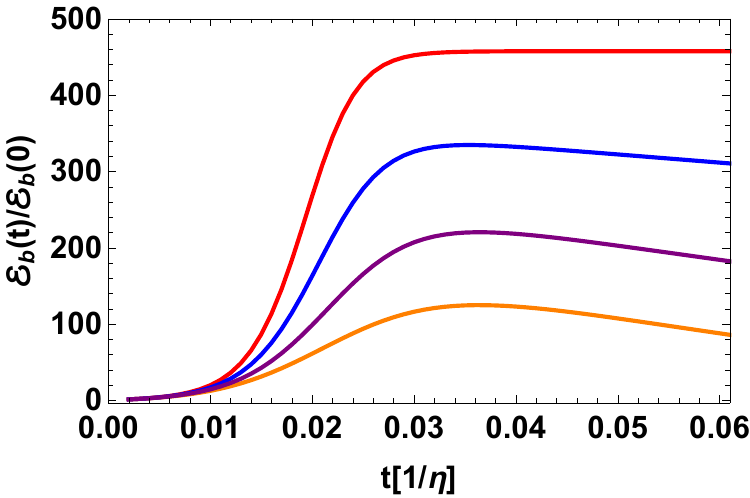}
	\end{center}
\caption{Evolution of $\x_\o$, cross helicity ${\cal H}_c$, and magnetic energy ${\cal E}_b$, normalized with their initial values. The lines in red, blue, purple, and orange correspond to $\eta\G=0, 0.01, 0.02$, and $0.03$, respectively.}
\label{fig:evo}
\end{figure}

When the chirality relaxation is negligible, $\G=0$, the coupled equations (\ref{eq:ch:cont2}), (\ref{fourier:eomz1}), and (\ref{fourier:eomz2}) permit a state in which $\m_5$, various helicities, and kinetic and magnetic energies are stationary. Such a state satisfies the condition that $\x_\o'=1$. To see this, we first observe from \eq{fourier:eomz1} that the magnetic diffusion would eventually diminish $z_{1\pm}$ and enforce $v_\pm = b'_\pm$ when $\x'_\o=1$. Then, \eq{ell:z2} and \eq{fourier:eomz2} gives that $v_\pm(t,\bk)=b'_\pm(t,\bk)\propto e^{i\bB'_0\cdot\bk t}$ representing a pure Alfv\'en wave (an Alfvenic state). In this case, ${\cal E}_{v,b}$, and ${\cal H}_{v,b,c}$ become time independent, which also implies the time independence of $\m_5$ according to \eq{eq:ch:cont2}. This suggests that when $\G=0$, the system will eventually evolve into a state such that $\x'_\o=1$, regardless of whether $\x'_\o$ is initially smaller or larger than $1$, as confirmed by numerical calculation given in Fig.~\ref{fig:evo}. When $\x'_\o$ is initially larger than $1$, this provides a dynamo mechanism. It has been long believed that the Alfvenic state is favored in relaxation processes in the MHD plasmas~\cite{Woltjer:1958}. Therefore, for a chiral plasma, the CMVI provides a mechanism for a fast reachment of such a Alfvenic state, in addition to other known effects~\cite{Matthaeus:2007pd}. Note that such a state maximizes the cross helicity ${\cal H}_c$ for a fixed total energy per unit mass ${\cal E}_v+{\cal E}_b$. 

When a finite $\G$ is present, $\m_5$ is constantly driven to zero. But this process can be very slow as usually $\G$ is small. As an illustration, in Fig.~\ref{fig:evo}, we show the time evolution of $\xi'_\o$, cross helicity ${\cal H}_c$, and magnetic energy ${\cal E}_b$, with an initial $\xi'_\o= 5$. To highlight the effect of CMVI, we have chosen an initial condition such that ${\cal H}_b={\cal H}_v=0$, which implies that they remain zero throughout the time evolution. Other parameters are chosen as follows: All the dimensionful quantities are in units of $1/\eta$, the backgroud magnetic field is $|\bB'_0|=5$, the initial $v_+(0)=b'_+(0)$ are given as a Fermi-Dirac shape $v_0/[\exp(10\eta |k_z|-100)+1]$ with $v_0=0.1$. It is evident from Fig.~\ref{fig:evo} that at the early times, the CMVI drives both the magnetic helicity and magnetic energy to grow exponentially. After that $\xi'_\o$ becomes smaller than $1$, the system evolves slowly (quasi-stationarily) towards vanishing velocity and magnetic field due to the finite $\G$. In such a way, the CMVI provides a fast dynamo mechanism by transferring chirality of constituent particles to the cross helicity of the system. It is worth noting that such a CMVI-induced dynamo mechanism bears some analogy with the turbulent cross helicity dynamo~\cite{Yokoi:2013mja,Yoshizawa:book}, in which the turbulent electromotive force $\lan\bv\times\bb\ran$ gains a term $\propto\bo$ due to the mean cross helicity in the plasma. This cross-helicity dynamo has been shown to play significant roles in geophysical and astrophysical plasmas~\cite{Yokoi:2013mja,Yoshizawa:book}. However, it is important to note that our CMVI-induced dynamo has a completely different origin from the cross-helicity dynamo, although they could act together in a turbulent chiral plasma.

{\bf Inclusion of chiral magnetic effect ---} In the above discussion, for the purpose of transparency, we have intentionally exluded the CME from the electric current. Upon restoring the CME, the linearized chiral MHD equations, expressed in terms of the reduced variables, become 
\begin{gather}
 \label{eq:mhd:ns4}
\pt_t\bv = \bB'_0\cdot\bnabla{\bm b}'\; ,\\
 \label{eq:mhd:mag4}
 \pt_t\bb'=\bB'_0\cdot\bnabla\bv+\eta\nabla^2\bb'+\eta\xi_B\bnabla\times\bb'-\eta\xi'_\o\nabla^2\bv.
\end{gather}
The dispersion relation of the eigenmodes of these equations are given by 
\begin{align} 
    \label{dispercvecme} 
    \o&=\o_\pm^\chi\equiv-\frac{i\eta}{2}(\bk^{2} -\c\xi_B|\bk|) \\
            &\pm \frac{1}{2}\sqrt{ 4(\bm{B}'_{0}\cdot \bm{k})^{2}-\eta^2(\bk^{2}-\c\xi_B|\bk|)^2- 4i \eta \xi'_\o  \bk^{2} \bm{B}'_{0}\cdot \bm{k} }\\  
     &\approx \pm\bB'_0\cdot\bk-i\frac{\eta}{2}\lb1\pm\xi'_\o\rb\bk^2 +i\c\frac{1}{2}\eta\x_B|\bk| +O(|\bk|^3),
\end{align}
where $\c=\pm$ corresponds to two different helicities in $\bb$. Therefore, we observe that, assuming $\x'_\o, \x_B>0$, when $\x'_\o>1$, there is always an unstable mode corresponding to $\o_-^{\c=+}$. Even when $\x'_\o<1$, the usual chiral plasma instability is catalyzed in a way that the modes with $|\bk|<\xi_B/(1-\x'_\o)$ are unstable, meaning that the unstable region in the wave number is enlarged from $|\bk|<\xi_B$~\cite{Joyce:1997uy,Akamatsu:2013pjd} to $|\bk|<\x_B/(1-\x'_\o)$. 

{\bf Discussion ---} To summarize, we have demonstrated that the presence of the CVE can induce a new type of plasma instability, the CMVI, in the presence of a background magnetic field. While the condition for CMVI to occur, $\x'_\o>1$, is stringent, we have discussed that other mechanisms, such as turbulence-induced cross helicity~\cite{Yokoi:2013mja,Yoshizawa:book}, may facilitate the onset of CMVI. Additionally, the combined effects of CVE and CME can broaden the kinematic region for the occurrence of chiral plasma instability, thus leaving a trace of CMVI. The CMVI can have interesting implications, e.g., it may lead to a new dynamo action and affect the evolution of the magnetic, cross, and kinetic helicities in chiral plasma. Possible applications include the electromagnetic plasma in astrophysical objects, the primordial electroweak plasma in early Univserse, the quark-gluon plasma in heavy-ion collisions, and the electron plasma in Dirac and Weyl semimetals.

{\bf Acknowledgments ---} This work is supported by the Natural Science Foundation of China (Grant No. 12147101, No. 12225502 and No. 12075061), the National Key Research and Development Program of China (Grant No. 2022YFA1604900), and the Natural Science Foundation of Shanghai (Grant No. 20ZR1404100).

\appendix
\section{Derivation of the MHD equations}\label{app:derivation}
For completeness, we provide a derivation of the chiral MHD equations (\ref{eq:mhd:ns}) and (\ref{eq:mhd:mag}) in this Appendix. A similar derivation for the conventional MHD equations can be found in textbooks such as ~\cite{Davidson:2017}. In the one-fluid description of the chiral plasma, the equation of motion for flow velocity $\bv$ is given by (where $P, \r, n$, and $\bj$ are pressure, mass density, charge density, and electric current, respectively)
\begin{gather}
 \label{eq:mhd:ns1}
\r\lb\pt_t +\bv\cdot\bm\nabla\rb\bv = -\bnabla P+n\bE+\bj\times\bB ,
\end{gather}
coupled with the Maxwell equations,
\begin{gather}
 \label{eq:mx:1}
\bnabla\cdot\bE=n ,\\
 \label{eq:mx:2}
\bnabla\cdot\bB=0 ,\\
 \label{eq:mx:3}
\bnabla\times\bB=\pt_t\bE+\bj ,\\
 \label{eq:mx:4}
\bnabla\times\bE=-\pt_t\bB ,
\end{gather}
and the continuity equations for $\r$ and $n$, 
\begin{gather}
 \label{eq:mhd:cont1}
\pt_t \r+\bnabla\cdot(\r\bv) =0 ,\\
 \label{eq:cons1}
\pt_t n +\bnabla\cdot\bj=0 .
\end{gather}
The constitutive relation for $\bj$ is 
\begin{align}
 \label{eq:app:const}
\bj =\bj_f+\bj_{\rm ohm}+\bj_B + \bj_\o,
\end{align}
with $\bj_f=n\bv$ the free current, $\bj_{\rm ohm}=\s(\bE+\bv\times\bB)$ the Ohmic current, $\bj_B=\xi_B\bB$ the CME current, and $\bj_\o=\xi_\o\bo$ the CVE current. We have omitted the viscous terms in \eq{eq:mhd:ns1} for the sake of simplicity. However, the effects of viscosity can be readily taken into account.

From \eq{eq:mx:4}, we have $|\bE|/|\bB|\sim L/\t\equiv u_0$ with $L, \t$ and $u_0$ the characteristic length, time, and velocity scales of the plasma. In the non-relativistic limit, $u_0\ll 1$, and one finds
\begin{gather}
 \label{eq:app:comp1}
\frac{|\pt_t\bE|}{|\bnabla\times\bB|}\sim u_0^2 \ll 1 ,\\
\frac{|n\bE|}{|\bj\times\bB|}\sim \frac{|\bnabla\cdot\bE||\bE|}{|(\bnabla\times\bB)\times\bB|} \sim u_0^2 \ll 1 ,\\
\frac{|\bj_f|}{|\bnabla\times\bB|}\sim \frac{|\bnabla\cdot\bE||\bv|}{|\bnabla\times\bB|} \sim u_0^2 \ll 1 .
\end{gather}
Therefore, we can eliminate $n\bE$ from \eq{eq:mhd:ns1}, $\pt_t\bE$ from \eq{eq:mx:3}, and $\bj_f$ from \eq{eq:app:const}. Consequently, the electric field $\bE$ is no longer a dynamical quantity and is determined by \eq{eq:app:const}, 
\begin{gather}
\bE=-\bv\times\bB +\eta(\bj-\bj_B-\bj_\o).
\end{gather}
Substituting \eq{eq:mx:3} into \eq{eq:mhd:ns1} and $\bE$ into \eq{eq:mx:4}, we obtain \eqs{eq:mhd:ns}{eq:mhd:mag} in the main text.


\bibliography{chiral_magvor_instability}

\begin{thebibliography}{55}%
\makeatletter
\providecommand \@ifxundefined [1]{%
 \@ifx{#1\undefined}
}%
\providecommand \@ifnum [1]{%
 \ifnum #1\expandafter \@firstoftwo
 \else \expandafter \@secondoftwo
 \fi
}%
\providecommand \@ifx [1]{%
 \ifx #1\expandafter \@firstoftwo
 \else \expandafter \@secondoftwo
 \fi
}%
\providecommand \natexlab [1]{#1}%
\providecommand \enquote  [1]{``#1''}%
\providecommand \bibnamefont  [1]{#1}%
\providecommand \bibfnamefont [1]{#1}%
\providecommand \citenamefont [1]{#1}%
\providecommand \href@noop [0]{\@secondoftwo}%
\providecommand \href [0]{\begingroup \@sanitize@url \@href}%
\providecommand \@href[1]{\@@startlink{#1}\@@href}%
\providecommand \@@href[1]{\endgroup#1\@@endlink}%
\providecommand \@sanitize@url [0]{\catcode `\\12\catcode `\$12\catcode
  `\&12\catcode `\#12\catcode `\^12\catcode `\_12\catcode `\%12\relax}%
\providecommand \@@startlink[1]{}%
\providecommand \@@endlink[0]{}%
\providecommand \url  [0]{\begingroup\@sanitize@url \@url }%
\providecommand \@url [1]{\endgroup\@href {#1}{\urlprefix }}%
\providecommand \urlprefix  [0]{URL }%
\providecommand \Eprint [0]{\href }%
\providecommand \doibase [0]{https://doi.org/}%
\providecommand \selectlanguage [0]{\@gobble}%
\providecommand \bibinfo  [0]{\@secondoftwo}%
\providecommand \bibfield  [0]{\@secondoftwo}%
\providecommand \translation [1]{[#1]}%
\providecommand \BibitemOpen [0]{}%
\providecommand \bibitemStop [0]{}%
\providecommand \bibitemNoStop [0]{.\EOS\space}%
\providecommand \EOS [0]{\spacefactor3000\relax}%
\providecommand \BibitemShut  [1]{\csname bibitem#1\endcsname}%
\let\auto@bib@innerbib\@empty
\bibitem [{\citenamefont {Vilenkin}(1980)}]{Vilenkin:1980fu}%
  \BibitemOpen
  \bibfield  {author} {\bibinfo {author} {\bibfnamefont {A.}~\bibnamefont
  {Vilenkin}},\ }\bibfield  {title} {\bibinfo {title} {{Equlibrium parity
  violating current in a magnetic field}},\ }\href
  {https://doi.org/10.1103/PhysRevD.22.3080} {\bibfield  {journal} {\bibinfo
  {journal} {Phys. Rev. D}\ }\textbf {\bibinfo {volume} {22}},\ \bibinfo
  {pages} {3080} (\bibinfo {year} {1980})}\BibitemShut {NoStop}%
\bibitem [{\citenamefont {Kharzeev}\ \emph {et~al.}(2008)\citenamefont
  {Kharzeev}, \citenamefont {McLerran},\ and\ \citenamefont
  {Warringa}}]{Kharzeev:2007jp}%
  \BibitemOpen
  \bibfield  {author} {\bibinfo {author} {\bibfnamefont {D.~E.}\ \bibnamefont
  {Kharzeev}}, \bibinfo {author} {\bibfnamefont {L.~D.}\ \bibnamefont
  {McLerran}},\ and\ \bibinfo {author} {\bibfnamefont {H.~J.}\ \bibnamefont
  {Warringa}},\ }\bibfield  {title} {\bibinfo {title} {{The Effects of
  topological charge change in heavy ion collisions: 'Event by event P and CP
  violation'}},\ }\href {https://doi.org/10.1016/j.nuclphysa.2008.02.298}
  {\bibfield  {journal} {\bibinfo  {journal} {Nucl. Phys. A}\ }\textbf
  {\bibinfo {volume} {803}},\ \bibinfo {pages} {227} (\bibinfo {year}
  {2008})},\ \Eprint {https://arxiv.org/abs/0711.0950} {arXiv:0711.0950
  [hep-ph]} \BibitemShut {NoStop}%
\bibitem [{\citenamefont {Fukushima}\ \emph {et~al.}(2008)\citenamefont
  {Fukushima}, \citenamefont {Kharzeev},\ and\ \citenamefont
  {Warringa}}]{Fukushima:2008xe}%
  \BibitemOpen
  \bibfield  {author} {\bibinfo {author} {\bibfnamefont {K.}~\bibnamefont
  {Fukushima}}, \bibinfo {author} {\bibfnamefont {D.~E.}\ \bibnamefont
  {Kharzeev}},\ and\ \bibinfo {author} {\bibfnamefont {H.~J.}\ \bibnamefont
  {Warringa}},\ }\bibfield  {title} {\bibinfo {title} {{The Chiral Magnetic
  Effect}},\ }\href {https://doi.org/10.1103/PhysRevD.78.074033} {\bibfield
  {journal} {\bibinfo  {journal} {Phys. Rev. D}\ }\textbf {\bibinfo {volume}
  {78}},\ \bibinfo {pages} {074033} (\bibinfo {year} {2008})},\ \Eprint
  {https://arxiv.org/abs/0808.3382} {arXiv:0808.3382 [hep-ph]} \BibitemShut
  {NoStop}%
\bibitem [{\citenamefont {Vilenkin}(1979)}]{Vilenkin:1979ui}%
  \BibitemOpen
  \bibfield  {author} {\bibinfo {author} {\bibfnamefont {A.}~\bibnamefont
  {Vilenkin}},\ }\bibfield  {title} {\bibinfo {title} {{Macroscopic parity
  violating effects: Neutrino fluxes from rotating black holes and in rotating
  thermal radiation}},\ }\href {https://doi.org/10.1103/PhysRevD.20.1807}
  {\bibfield  {journal} {\bibinfo  {journal} {Phys. Rev. D}\ }\textbf {\bibinfo
  {volume} {20}},\ \bibinfo {pages} {1807} (\bibinfo {year}
  {1979})}\BibitemShut {NoStop}%
\bibitem [{\citenamefont {Erdmenger}\ \emph {et~al.}(2009)\citenamefont
  {Erdmenger}, \citenamefont {Haack}, \citenamefont {Kaminski},\ and\
  \citenamefont {Yarom}}]{Erdmenger:2008rm}%
  \BibitemOpen
  \bibfield  {author} {\bibinfo {author} {\bibfnamefont {J.}~\bibnamefont
  {Erdmenger}}, \bibinfo {author} {\bibfnamefont {M.}~\bibnamefont {Haack}},
  \bibinfo {author} {\bibfnamefont {M.}~\bibnamefont {Kaminski}},\ and\
  \bibinfo {author} {\bibfnamefont {A.}~\bibnamefont {Yarom}},\ }\bibfield
  {title} {\bibinfo {title} {{Fluid dynamics of R-charged black holes}},\
  }\href {https://doi.org/10.1088/1126-6708/2009/01/055} {\bibfield  {journal}
  {\bibinfo  {journal} {JHEP}\ }\textbf {\bibinfo {volume} {01}},\ \bibinfo
  {pages} {055}},\ \Eprint {https://arxiv.org/abs/0809.2488} {arXiv:0809.2488
  [hep-th]} \BibitemShut {NoStop}%
\bibitem [{\citenamefont {Banerjee}\ \emph {et~al.}(2011)\citenamefont
  {Banerjee}, \citenamefont {Bhattacharya}, \citenamefont {Bhattacharyya},
  \citenamefont {Dutta}, \citenamefont {Loganayagam},\ and\ \citenamefont
  {Surowka}}]{Banerjee:2008th}%
  \BibitemOpen
  \bibfield  {author} {\bibinfo {author} {\bibfnamefont {N.}~\bibnamefont
  {Banerjee}}, \bibinfo {author} {\bibfnamefont {J.}~\bibnamefont
  {Bhattacharya}}, \bibinfo {author} {\bibfnamefont {S.}~\bibnamefont
  {Bhattacharyya}}, \bibinfo {author} {\bibfnamefont {S.}~\bibnamefont
  {Dutta}}, \bibinfo {author} {\bibfnamefont {R.}~\bibnamefont {Loganayagam}},\
  and\ \bibinfo {author} {\bibfnamefont {P.}~\bibnamefont {Surowka}},\
  }\bibfield  {title} {\bibinfo {title} {{Hydrodynamics from charged black
  branes}},\ }\href {https://doi.org/10.1007/JHEP01(2011)094} {\bibfield
  {journal} {\bibinfo  {journal} {JHEP}\ }\textbf {\bibinfo {volume} {01}},\
  \bibinfo {pages} {094}},\ \Eprint {https://arxiv.org/abs/0809.2596}
  {arXiv:0809.2596 [hep-th]} \BibitemShut {NoStop}%
\bibitem [{\citenamefont {Son}\ and\ \citenamefont
  {Surowka}(2009)}]{Son:2009tf}%
  \BibitemOpen
  \bibfield  {author} {\bibinfo {author} {\bibfnamefont {D.~T.}\ \bibnamefont
  {Son}}\ and\ \bibinfo {author} {\bibfnamefont {P.}~\bibnamefont {Surowka}},\
  }\bibfield  {title} {\bibinfo {title} {{Hydrodynamics with Triangle
  Anomalies}},\ }\href {https://doi.org/10.1103/PhysRevLett.103.191601}
  {\bibfield  {journal} {\bibinfo  {journal} {Phys. Rev. Lett.}\ }\textbf
  {\bibinfo {volume} {103}},\ \bibinfo {pages} {191601} (\bibinfo {year}
  {2009})},\ \Eprint {https://arxiv.org/abs/0906.5044} {arXiv:0906.5044
  [hep-th]} \BibitemShut {NoStop}%
\bibitem [{\citenamefont {Huang}(2016)}]{Huang:2015oca}%
  \BibitemOpen
  \bibfield  {author} {\bibinfo {author} {\bibfnamefont {X.-G.}\ \bibnamefont
  {Huang}},\ }\bibfield  {title} {\bibinfo {title} {{Electromagnetic fields and
  anomalous transports in heavy-ion collisions --- A pedagogical review}},\
  }\href {https://doi.org/10.1088/0034-4885/79/7/076302} {\bibfield  {journal}
  {\bibinfo  {journal} {Rept. Prog. Phys.}\ }\textbf {\bibinfo {volume} {79}},\
  \bibinfo {pages} {076302} (\bibinfo {year} {2016})},\ \Eprint
  {https://arxiv.org/abs/1509.04073} {arXiv:1509.04073 [nucl-th]} \BibitemShut
  {NoStop}%
\bibitem [{\citenamefont {Liu}\ and\ \citenamefont
  {Huang}(2020)}]{Liu:2020ymh}%
  \BibitemOpen
  \bibfield  {author} {\bibinfo {author} {\bibfnamefont {Y.-C.}\ \bibnamefont
  {Liu}}\ and\ \bibinfo {author} {\bibfnamefont {X.-G.}\ \bibnamefont
  {Huang}},\ }\bibfield  {title} {\bibinfo {title} {{Anomalous chiral
  transports and spin polarization in heavy-ion collisions}},\ }\href
  {https://doi.org/10.1007/s41365-020-00764-z} {\bibfield  {journal} {\bibinfo
  {journal} {Nucl. Sci. Tech.}\ }\textbf {\bibinfo {volume} {31}},\ \bibinfo
  {pages} {56} (\bibinfo {year} {2020})},\ \Eprint
  {https://arxiv.org/abs/2003.12482} {arXiv:2003.12482 [nucl-th]} \BibitemShut
  {NoStop}%
\bibitem [{\citenamefont {Kharzeev}\ and\ \citenamefont
  {Liao}(2021)}]{Kharzeev:2020jxw}%
  \BibitemOpen
  \bibfield  {author} {\bibinfo {author} {\bibfnamefont {D.~E.}\ \bibnamefont
  {Kharzeev}}\ and\ \bibinfo {author} {\bibfnamefont {J.}~\bibnamefont
  {Liao}},\ }\bibfield  {title} {\bibinfo {title} {{Chiral magnetic effect
  reveals the topology of gauge fields in heavy-ion collisions}},\ }\href
  {https://doi.org/10.1038/s42254-020-00254-6} {\bibfield  {journal} {\bibinfo
  {journal} {Nature Rev. Phys.}\ }\textbf {\bibinfo {volume} {3}},\ \bibinfo
  {pages} {55} (\bibinfo {year} {2021})},\ \Eprint
  {https://arxiv.org/abs/2102.06623} {arXiv:2102.06623 [hep-ph]} \BibitemShut
  {NoStop}%
\bibitem [{\citenamefont {Chernodub}\ \emph {et~al.}(2022)\citenamefont
  {Chernodub}, \citenamefont {Ferreiros}, \citenamefont {Grushin},
  \citenamefont {Landsteiner},\ and\ \citenamefont
  {Vozmediano}}]{Chernodub:2021nff}%
  \BibitemOpen
  \bibfield  {author} {\bibinfo {author} {\bibfnamefont {M.~N.}\ \bibnamefont
  {Chernodub}}, \bibinfo {author} {\bibfnamefont {Y.}~\bibnamefont
  {Ferreiros}}, \bibinfo {author} {\bibfnamefont {A.~G.}\ \bibnamefont
  {Grushin}}, \bibinfo {author} {\bibfnamefont {K.}~\bibnamefont
  {Landsteiner}},\ and\ \bibinfo {author} {\bibfnamefont {M.~A.~H.}\
  \bibnamefont {Vozmediano}},\ }\bibfield  {title} {\bibinfo {title} {{Thermal
  transport, geometry, and anomalies}},\ }\href
  {https://doi.org/10.1016/j.physrep.2022.06.002} {\bibfield  {journal}
  {\bibinfo  {journal} {Phys. Rept.}\ }\textbf {\bibinfo {volume} {977}},\
  \bibinfo {pages} {1} (\bibinfo {year} {2022})},\ \Eprint
  {https://arxiv.org/abs/2110.05471} {arXiv:2110.05471 [cond-mat.mes-hall]}
  \BibitemShut {NoStop}%
\bibitem [{\citenamefont {Kharzeev}\ and\ \citenamefont
  {Yee}(2011)}]{Kharzeev:2010gd}%
  \BibitemOpen
  \bibfield  {author} {\bibinfo {author} {\bibfnamefont {D.~E.}\ \bibnamefont
  {Kharzeev}}\ and\ \bibinfo {author} {\bibfnamefont {H.-U.}\ \bibnamefont
  {Yee}},\ }\bibfield  {title} {\bibinfo {title} {{Chiral Magnetic Wave}},\
  }\href {https://doi.org/10.1103/PhysRevD.83.085007} {\bibfield  {journal}
  {\bibinfo  {journal} {Phys. Rev. D}\ }\textbf {\bibinfo {volume} {83}},\
  \bibinfo {pages} {085007} (\bibinfo {year} {2011})},\ \Eprint
  {https://arxiv.org/abs/1012.6026} {arXiv:1012.6026 [hep-th]} \BibitemShut
  {NoStop}%
\bibitem [{\citenamefont {Jiang}\ \emph {et~al.}(2015)\citenamefont {Jiang},
  \citenamefont {Huang},\ and\ \citenamefont {Liao}}]{Jiang:2015cva}%
  \BibitemOpen
  \bibfield  {author} {\bibinfo {author} {\bibfnamefont {Y.}~\bibnamefont
  {Jiang}}, \bibinfo {author} {\bibfnamefont {X.-G.}\ \bibnamefont {Huang}},\
  and\ \bibinfo {author} {\bibfnamefont {J.}~\bibnamefont {Liao}},\ }\bibfield
  {title} {\bibinfo {title} {{Chiral vortical wave and induced flavor charge
  transport in a rotating quark-gluon plasma}},\ }\href
  {https://doi.org/10.1103/PhysRevD.92.071501} {\bibfield  {journal} {\bibinfo
  {journal} {Phys. Rev. D}\ }\textbf {\bibinfo {volume} {92}},\ \bibinfo
  {pages} {071501} (\bibinfo {year} {2015})},\ \Eprint
  {https://arxiv.org/abs/1504.03201} {arXiv:1504.03201 [hep-ph]} \BibitemShut
  {NoStop}%
\bibitem [{\citenamefont {Huang}\ and\ \citenamefont
  {Liao}(2013)}]{Huang:2013iia}%
  \BibitemOpen
  \bibfield  {author} {\bibinfo {author} {\bibfnamefont {X.-G.}\ \bibnamefont
  {Huang}}\ and\ \bibinfo {author} {\bibfnamefont {J.}~\bibnamefont {Liao}},\
  }\bibfield  {title} {\bibinfo {title} {{Axial Current Generation from
  Electric Field: Chiral Electric Separation Effect}},\ }\href
  {https://doi.org/10.1103/PhysRevLett.110.232302} {\bibfield  {journal}
  {\bibinfo  {journal} {Phys. Rev. Lett.}\ }\textbf {\bibinfo {volume} {110}},\
  \bibinfo {pages} {232302} (\bibinfo {year} {2013})},\ \Eprint
  {https://arxiv.org/abs/1303.7192} {arXiv:1303.7192 [nucl-th]} \BibitemShut
  {NoStop}%
\bibitem [{\citenamefont {Yamamoto}(2015)}]{Yamamoto:2015ria}%
  \BibitemOpen
  \bibfield  {author} {\bibinfo {author} {\bibfnamefont {N.}~\bibnamefont
  {Yamamoto}},\ }\bibfield  {title} {\bibinfo {title} {{Chiral Alfv\'en Wave in
  Anomalous Hydrodynamics}},\ }\href
  {https://doi.org/10.1103/PhysRevLett.115.141601} {\bibfield  {journal}
  {\bibinfo  {journal} {Phys. Rev. Lett.}\ }\textbf {\bibinfo {volume} {115}},\
  \bibinfo {pages} {141601} (\bibinfo {year} {2015})},\ \Eprint
  {https://arxiv.org/abs/1505.05444} {arXiv:1505.05444 [hep-th]} \BibitemShut
  {NoStop}%
\bibitem [{\citenamefont {Chernodub}(2016)}]{Chernodub:2015gxa}%
  \BibitemOpen
  \bibfield  {author} {\bibinfo {author} {\bibfnamefont {M.~N.}\ \bibnamefont
  {Chernodub}},\ }\bibfield  {title} {\bibinfo {title} {{Chiral Heat Wave and
  mixing of Magnetic, Vortical and Heat waves in chiral media}},\ }\href
  {https://doi.org/10.1007/JHEP01(2016)100} {\bibfield  {journal} {\bibinfo
  {journal} {JHEP}\ }\textbf {\bibinfo {volume} {01}},\ \bibinfo {pages}
  {100}},\ \Eprint {https://arxiv.org/abs/1509.01245} {arXiv:1509.01245
  [hep-th]} \BibitemShut {NoStop}%
\bibitem [{\citenamefont {Joyce}\ and\ \citenamefont
  {Shaposhnikov}(1997)}]{Joyce:1997uy}%
  \BibitemOpen
  \bibfield  {author} {\bibinfo {author} {\bibfnamefont {M.}~\bibnamefont
  {Joyce}}\ and\ \bibinfo {author} {\bibfnamefont {M.~E.}\ \bibnamefont
  {Shaposhnikov}},\ }\bibfield  {title} {\bibinfo {title} {{Primordial magnetic
  fields, right-handed electrons, and the Abelian anomaly}},\ }\href
  {https://doi.org/10.1103/PhysRevLett.79.1193} {\bibfield  {journal} {\bibinfo
   {journal} {Phys. Rev. Lett.}\ }\textbf {\bibinfo {volume} {79}},\ \bibinfo
  {pages} {1193} (\bibinfo {year} {1997})},\ \Eprint
  {https://arxiv.org/abs/astro-ph/9703005} {arXiv:astro-ph/9703005}
  \BibitemShut {NoStop}%
\bibitem [{\citenamefont {Akamatsu}\ and\ \citenamefont
  {Yamamoto}(2013)}]{Akamatsu:2013pjd}%
  \BibitemOpen
  \bibfield  {author} {\bibinfo {author} {\bibfnamefont {Y.}~\bibnamefont
  {Akamatsu}}\ and\ \bibinfo {author} {\bibfnamefont {N.}~\bibnamefont
  {Yamamoto}},\ }\bibfield  {title} {\bibinfo {title} {{Chiral Plasma
  Instabilities}},\ }\href {https://doi.org/10.1103/PhysRevLett.111.052002}
  {\bibfield  {journal} {\bibinfo  {journal} {Phys. Rev. Lett.}\ }\textbf
  {\bibinfo {volume} {111}},\ \bibinfo {pages} {052002} (\bibinfo {year}
  {2013})},\ \Eprint {https://arxiv.org/abs/1302.2125} {arXiv:1302.2125
  [nucl-th]} \BibitemShut {NoStop}%
\bibitem [{\citenamefont {Giovannini}\ and\ \citenamefont
  {Shaposhnikov}(1998)}]{Giovannini:1997eg}%
  \BibitemOpen
  \bibfield  {author} {\bibinfo {author} {\bibfnamefont {M.}~\bibnamefont
  {Giovannini}}\ and\ \bibinfo {author} {\bibfnamefont {M.~E.}\ \bibnamefont
  {Shaposhnikov}},\ }\bibfield  {title} {\bibinfo {title} {{Primordial
  hypermagnetic fields and triangle anomaly}},\ }\href
  {https://doi.org/10.1103/PhysRevD.57.2186} {\bibfield  {journal} {\bibinfo
  {journal} {Phys. Rev. D}\ }\textbf {\bibinfo {volume} {57}},\ \bibinfo
  {pages} {2186} (\bibinfo {year} {1998})},\ \Eprint
  {https://arxiv.org/abs/hep-ph/9710234} {arXiv:hep-ph/9710234} \BibitemShut
  {NoStop}%
\bibitem [{\citenamefont {Boyarsky}\ \emph
  {et~al.}(2012{\natexlab{a}})\citenamefont {Boyarsky}, \citenamefont
  {Frohlich},\ and\ \citenamefont {Ruchayskiy}}]{Boyarsky:2011uy}%
  \BibitemOpen
  \bibfield  {author} {\bibinfo {author} {\bibfnamefont {A.}~\bibnamefont
  {Boyarsky}}, \bibinfo {author} {\bibfnamefont {J.}~\bibnamefont {Frohlich}},\
  and\ \bibinfo {author} {\bibfnamefont {O.}~\bibnamefont {Ruchayskiy}},\
  }\bibfield  {title} {\bibinfo {title} {{Self-consistent evolution of magnetic
  fields and chiral asymmetry in the early Universe}},\ }\href
  {https://doi.org/10.1103/PhysRevLett.108.031301} {\bibfield  {journal}
  {\bibinfo  {journal} {Phys. Rev. Lett.}\ }\textbf {\bibinfo {volume} {108}},\
  \bibinfo {pages} {031301} (\bibinfo {year} {2012}{\natexlab{a}})},\ \Eprint
  {https://arxiv.org/abs/1109.3350} {arXiv:1109.3350 [astro-ph.CO]}
  \BibitemShut {NoStop}%
\bibitem [{\citenamefont {Boyarsky}\ \emph
  {et~al.}(2012{\natexlab{b}})\citenamefont {Boyarsky}, \citenamefont
  {Ruchayskiy},\ and\ \citenamefont {Shaposhnikov}}]{Boyarsky:2012ex}%
  \BibitemOpen
  \bibfield  {author} {\bibinfo {author} {\bibfnamefont {A.}~\bibnamefont
  {Boyarsky}}, \bibinfo {author} {\bibfnamefont {O.}~\bibnamefont
  {Ruchayskiy}},\ and\ \bibinfo {author} {\bibfnamefont {M.}~\bibnamefont
  {Shaposhnikov}},\ }\bibfield  {title} {\bibinfo {title} {{Long-range magnetic
  fields in the ground state of the Standard Model plasma}},\ }\href
  {https://doi.org/10.1103/PhysRevLett.109.111602} {\bibfield  {journal}
  {\bibinfo  {journal} {Phys. Rev. Lett.}\ }\textbf {\bibinfo {volume} {109}},\
  \bibinfo {pages} {111602} (\bibinfo {year} {2012}{\natexlab{b}})},\ \Eprint
  {https://arxiv.org/abs/1204.3604} {arXiv:1204.3604 [hep-ph]} \BibitemShut
  {NoStop}%
\bibitem [{\citenamefont {Tashiro}\ \emph {et~al.}(2012)\citenamefont
  {Tashiro}, \citenamefont {Vachaspati},\ and\ \citenamefont
  {Vilenkin}}]{Tashiro:2012mf}%
  \BibitemOpen
  \bibfield  {author} {\bibinfo {author} {\bibfnamefont {H.}~\bibnamefont
  {Tashiro}}, \bibinfo {author} {\bibfnamefont {T.}~\bibnamefont
  {Vachaspati}},\ and\ \bibinfo {author} {\bibfnamefont {A.}~\bibnamefont
  {Vilenkin}},\ }\bibfield  {title} {\bibinfo {title} {{Chiral Effects and
  Cosmic Magnetic Fields}},\ }\href
  {https://doi.org/10.1103/PhysRevD.86.105033} {\bibfield  {journal} {\bibinfo
  {journal} {Phys. Rev. D}\ }\textbf {\bibinfo {volume} {86}},\ \bibinfo
  {pages} {105033} (\bibinfo {year} {2012})},\ \Eprint
  {https://arxiv.org/abs/1206.5549} {arXiv:1206.5549 [astro-ph.CO]}
  \BibitemShut {NoStop}%
\bibitem [{\citenamefont {Dvornikov}\ and\ \citenamefont
  {Semikoz}(2017)}]{Dvornikov:2016jth}%
  \BibitemOpen
  \bibfield  {author} {\bibinfo {author} {\bibfnamefont {M.}~\bibnamefont
  {Dvornikov}}\ and\ \bibinfo {author} {\bibfnamefont {V.~B.}\ \bibnamefont
  {Semikoz}},\ }\bibfield  {title} {\bibinfo {title} {{Influence of the
  turbulent motion on the chiral magnetic effect in the early Universe}},\
  }\href {https://doi.org/10.1103/PhysRevD.95.043538} {\bibfield  {journal}
  {\bibinfo  {journal} {Phys. Rev. D}\ }\textbf {\bibinfo {volume} {95}},\
  \bibinfo {pages} {043538} (\bibinfo {year} {2017})},\ \Eprint
  {https://arxiv.org/abs/1612.05897} {arXiv:1612.05897 [astro-ph.CO]}
  \BibitemShut {NoStop}%
\bibitem [{\citenamefont {Gorbar}\ \emph {et~al.}(2016)\citenamefont {Gorbar},
  \citenamefont {Rudenok}, \citenamefont {Shovkovy},\ and\ \citenamefont
  {Vilchinskii}}]{Gorbar:2016klv}%
  \BibitemOpen
  \bibfield  {author} {\bibinfo {author} {\bibfnamefont {E.~V.}\ \bibnamefont
  {Gorbar}}, \bibinfo {author} {\bibfnamefont {I.}~\bibnamefont {Rudenok}},
  \bibinfo {author} {\bibfnamefont {I.~A.}\ \bibnamefont {Shovkovy}},\ and\
  \bibinfo {author} {\bibfnamefont {S.}~\bibnamefont {Vilchinskii}},\
  }\bibfield  {title} {\bibinfo {title} {{Anomaly-driven inverse cascade and
  inhomogeneities in a magnetized chiral plasma in the early Universe}},\
  }\href {https://doi.org/10.1103/PhysRevD.94.103528} {\bibfield  {journal}
  {\bibinfo  {journal} {Phys. Rev. D}\ }\textbf {\bibinfo {volume} {94}},\
  \bibinfo {pages} {103528} (\bibinfo {year} {2016})},\ \Eprint
  {https://arxiv.org/abs/1610.01214} {arXiv:1610.01214 [hep-ph]} \BibitemShut
  {NoStop}%
\bibitem [{\citenamefont {Brandenburg}\ \emph {et~al.}(2017)\citenamefont
  {Brandenburg}, \citenamefont {Schober}, \citenamefont {Rogachevskii},
  \citenamefont {Kahniashvili}, \citenamefont {Boyarsky}, \citenamefont
  {Frohlich}, \citenamefont {Ruchayskiy},\ and\ \citenamefont
  {Kleeorin}}]{Brandenburg:2017rcb}%
  \BibitemOpen
  \bibfield  {author} {\bibinfo {author} {\bibfnamefont {A.}~\bibnamefont
  {Brandenburg}}, \bibinfo {author} {\bibfnamefont {J.}~\bibnamefont
  {Schober}}, \bibinfo {author} {\bibfnamefont {I.}~\bibnamefont
  {Rogachevskii}}, \bibinfo {author} {\bibfnamefont {T.}~\bibnamefont
  {Kahniashvili}}, \bibinfo {author} {\bibfnamefont {A.}~\bibnamefont
  {Boyarsky}}, \bibinfo {author} {\bibfnamefont {J.}~\bibnamefont {Frohlich}},
  \bibinfo {author} {\bibfnamefont {O.}~\bibnamefont {Ruchayskiy}},\ and\
  \bibinfo {author} {\bibfnamefont {N.}~\bibnamefont {Kleeorin}},\ }\bibfield
  {title} {\bibinfo {title} {{The turbulent chiral-magnetic cascade in the
  early universe}},\ }\href {https://doi.org/10.3847/2041-8213/aa855d}
  {\bibfield  {journal} {\bibinfo  {journal} {Astrophys. J. Lett.}\ }\textbf
  {\bibinfo {volume} {845}},\ \bibinfo {pages} {L21} (\bibinfo {year}
  {2017})},\ \Eprint {https://arxiv.org/abs/1707.03385} {arXiv:1707.03385
  [astro-ph.CO]} \BibitemShut {NoStop}%
\bibitem [{\citenamefont {Schober}\ \emph {et~al.}(2018)\citenamefont
  {Schober}, \citenamefont {Rogachevskii}, \citenamefont {Brandenburg},
  \citenamefont {Boyarsky}, \citenamefont {Fr\"ohlich}, \citenamefont
  {Ruchayskiy},\ and\ \citenamefont {Kleeorin}}]{Schober:2017cdw}%
  \BibitemOpen
  \bibfield  {author} {\bibinfo {author} {\bibfnamefont {J.}~\bibnamefont
  {Schober}}, \bibinfo {author} {\bibfnamefont {I.}~\bibnamefont
  {Rogachevskii}}, \bibinfo {author} {\bibfnamefont {A.}~\bibnamefont
  {Brandenburg}}, \bibinfo {author} {\bibfnamefont {A.}~\bibnamefont
  {Boyarsky}}, \bibinfo {author} {\bibfnamefont {J.}~\bibnamefont
  {Fr\"ohlich}}, \bibinfo {author} {\bibfnamefont {O.}~\bibnamefont
  {Ruchayskiy}},\ and\ \bibinfo {author} {\bibfnamefont {N.}~\bibnamefont
  {Kleeorin}},\ }\bibfield  {title} {\bibinfo {title} {{Laminar and turbulent
  dynamos in chiral magnetohydrodynamics. II. Simulations}},\ }\href
  {https://doi.org/10.3847/1538-4357/aaba75} {\bibfield  {journal} {\bibinfo
  {journal} {Astrophys. J.}\ }\textbf {\bibinfo {volume} {858}},\ \bibinfo
  {pages} {124} (\bibinfo {year} {2018})},\ \Eprint
  {https://arxiv.org/abs/1711.09733} {arXiv:1711.09733 [physics.flu-dyn]}
  \BibitemShut {NoStop}%
\bibitem [{\citenamefont {Ohnishi}\ and\ \citenamefont
  {Yamamoto}(2014)}]{Ohnishi:2014uea}%
  \BibitemOpen
  \bibfield  {author} {\bibinfo {author} {\bibfnamefont {A.}~\bibnamefont
  {Ohnishi}}\ and\ \bibinfo {author} {\bibfnamefont {N.}~\bibnamefont
  {Yamamoto}},\ }\bibfield  {title} {\bibinfo {title} {{Magnetars and the
  Chiral Plasma Instabilities}},\ }\href@noop {} {\  (\bibinfo {year}
  {2014})},\ \Eprint {https://arxiv.org/abs/1402.4760} {arXiv:1402.4760
  [astro-ph.HE]} \BibitemShut {NoStop}%
\bibitem [{\citenamefont {Yamamoto}(2016)}]{Yamamoto:2015gzz}%
  \BibitemOpen
  \bibfield  {author} {\bibinfo {author} {\bibfnamefont {N.}~\bibnamefont
  {Yamamoto}},\ }\bibfield  {title} {\bibinfo {title} {{Chiral transport of
  neutrinos in supernovae: Neutrino-induced fluid helicity and helical plasma
  instability}},\ }\href {https://doi.org/10.1103/PhysRevD.93.065017}
  {\bibfield  {journal} {\bibinfo  {journal} {Phys. Rev. D}\ }\textbf {\bibinfo
  {volume} {93}},\ \bibinfo {pages} {065017} (\bibinfo {year} {2016})},\
  \Eprint {https://arxiv.org/abs/1511.00933} {arXiv:1511.00933 [astro-ph.HE]}
  \BibitemShut {NoStop}%
\bibitem [{\citenamefont {Grabowska}\ \emph {et~al.}(2015)\citenamefont
  {Grabowska}, \citenamefont {Kaplan},\ and\ \citenamefont
  {Reddy}}]{Grabowska:2014efa}%
  \BibitemOpen
  \bibfield  {author} {\bibinfo {author} {\bibfnamefont {D.}~\bibnamefont
  {Grabowska}}, \bibinfo {author} {\bibfnamefont {D.~B.}\ \bibnamefont
  {Kaplan}},\ and\ \bibinfo {author} {\bibfnamefont {S.}~\bibnamefont
  {Reddy}},\ }\bibfield  {title} {\bibinfo {title} {{Role of the electron mass
  in damping chiral plasma instability in Supernovae and neutron stars}},\
  }\href {https://doi.org/10.1103/PhysRevD.91.085035} {\bibfield  {journal}
  {\bibinfo  {journal} {Phys. Rev. D}\ }\textbf {\bibinfo {volume} {91}},\
  \bibinfo {pages} {085035} (\bibinfo {year} {2015})},\ \Eprint
  {https://arxiv.org/abs/1409.3602} {arXiv:1409.3602 [hep-ph]} \BibitemShut
  {NoStop}%
\bibitem [{\citenamefont {Dvornikov}\ and\ \citenamefont
  {Semikoz}(2015)}]{Dvornikov:2014uza}%
  \BibitemOpen
  \bibfield  {author} {\bibinfo {author} {\bibfnamefont {M.}~\bibnamefont
  {Dvornikov}}\ and\ \bibinfo {author} {\bibfnamefont {V.~B.}\ \bibnamefont
  {Semikoz}},\ }\bibfield  {title} {\bibinfo {title} {{Magnetic field
  instability in a neutron star driven by the electroweak electron-nucleon
  interaction versus the chiral magnetic effect}},\ }\href
  {https://doi.org/10.1103/PhysRevD.91.061301} {\bibfield  {journal} {\bibinfo
  {journal} {Phys. Rev. D}\ }\textbf {\bibinfo {volume} {91}},\ \bibinfo
  {pages} {061301} (\bibinfo {year} {2015})},\ \Eprint
  {https://arxiv.org/abs/1410.6676} {arXiv:1410.6676 [astro-ph.HE]}
  \BibitemShut {NoStop}%
\bibitem [{\citenamefont {Sigl}\ and\ \citenamefont
  {Leite}(2016)}]{Sigl:2015xva}%
  \BibitemOpen
  \bibfield  {author} {\bibinfo {author} {\bibfnamefont {G.}~\bibnamefont
  {Sigl}}\ and\ \bibinfo {author} {\bibfnamefont {N.}~\bibnamefont {Leite}},\
  }\bibfield  {title} {\bibinfo {title} {{Chiral Magnetic Effect in
  Protoneutron Stars and Magnetic Field Spectral Evolution}},\ }\href
  {https://doi.org/10.1088/1475-7516/2016/01/025} {\bibfield  {journal}
  {\bibinfo  {journal} {JCAP}\ }\textbf {\bibinfo {volume} {01}},\ \bibinfo
  {pages} {025}},\ \Eprint {https://arxiv.org/abs/1507.04983} {arXiv:1507.04983
  [astro-ph.HE]} \BibitemShut {NoStop}%
\bibitem [{\citenamefont {Matsumoto}\ \emph {et~al.}(2022)\citenamefont
  {Matsumoto}, \citenamefont {Yamamoto},\ and\ \citenamefont
  {Yang}}]{Matsumoto:2022lyb}%
  \BibitemOpen
  \bibfield  {author} {\bibinfo {author} {\bibfnamefont {J.}~\bibnamefont
  {Matsumoto}}, \bibinfo {author} {\bibfnamefont {N.}~\bibnamefont
  {Yamamoto}},\ and\ \bibinfo {author} {\bibfnamefont {D.-L.}\ \bibnamefont
  {Yang}},\ }\bibfield  {title} {\bibinfo {title} {{Chiral plasma instability
  and inverse cascade from nonequilibrium left-handed neutrinos in
  core-collapse supernovae}},\ }\href
  {https://doi.org/10.1103/PhysRevD.105.123029} {\bibfield  {journal} {\bibinfo
   {journal} {Phys. Rev. D}\ }\textbf {\bibinfo {volume} {105}},\ \bibinfo
  {pages} {123029} (\bibinfo {year} {2022})},\ \Eprint
  {https://arxiv.org/abs/2202.09205} {arXiv:2202.09205 [astro-ph.HE]}
  \BibitemShut {NoStop}%
\bibitem [{\citenamefont {Schober}\ \emph {et~al.}(2022)\citenamefont
  {Schober}, \citenamefont {Rogachevskii},\ and\ \citenamefont
  {Brandenburg}}]{Schober:2021yav}%
  \BibitemOpen
  \bibfield  {author} {\bibinfo {author} {\bibfnamefont {J.}~\bibnamefont
  {Schober}}, \bibinfo {author} {\bibfnamefont {I.}~\bibnamefont
  {Rogachevskii}},\ and\ \bibinfo {author} {\bibfnamefont {A.}~\bibnamefont
  {Brandenburg}},\ }\bibfield  {title} {\bibinfo {title} {{Production of a
  Chiral Magnetic Anomaly with Emerging Turbulence and Mean-Field Dynamo
  Action}},\ }\href {https://doi.org/10.1103/PhysRevLett.128.065002} {\bibfield
   {journal} {\bibinfo  {journal} {Phys. Rev. Lett.}\ }\textbf {\bibinfo
  {volume} {128}},\ \bibinfo {pages} {065002} (\bibinfo {year} {2022})},\
  \Eprint {https://arxiv.org/abs/2107.12945} {arXiv:2107.12945
  [physics.plasm-ph]} \BibitemShut {NoStop}%
\bibitem [{\citenamefont {Brandenburg}\ \emph {et~al.}(2023)\citenamefont
  {Brandenburg}, \citenamefont {Kamada}, \citenamefont {Mukaida}, \citenamefont
  {Schmitz},\ and\ \citenamefont {Schober}}]{Brandenburg:2023aco}%
  \BibitemOpen
  \bibfield  {author} {\bibinfo {author} {\bibfnamefont {A.}~\bibnamefont
  {Brandenburg}}, \bibinfo {author} {\bibfnamefont {K.}~\bibnamefont {Kamada}},
  \bibinfo {author} {\bibfnamefont {K.}~\bibnamefont {Mukaida}}, \bibinfo
  {author} {\bibfnamefont {K.}~\bibnamefont {Schmitz}},\ and\ \bibinfo {author}
  {\bibfnamefont {J.}~\bibnamefont {Schober}},\ }\bibfield  {title} {\bibinfo
  {title} {{Chiral magnetohydrodynamics with zero total chirality}},\
  }\href@noop {} {\  (\bibinfo {year} {2023})},\ \Eprint
  {https://arxiv.org/abs/2304.06612} {arXiv:2304.06612 [hep-ph]} \BibitemShut
  {NoStop}%
\bibitem [{\citenamefont {Akamatsu}\ \emph {et~al.}(2016)\citenamefont
  {Akamatsu}, \citenamefont {Rothkopf},\ and\ \citenamefont
  {Yamamoto}}]{Akamatsu:2015kau}%
  \BibitemOpen
  \bibfield  {author} {\bibinfo {author} {\bibfnamefont {Y.}~\bibnamefont
  {Akamatsu}}, \bibinfo {author} {\bibfnamefont {A.}~\bibnamefont {Rothkopf}},\
  and\ \bibinfo {author} {\bibfnamefont {N.}~\bibnamefont {Yamamoto}},\
  }\bibfield  {title} {\bibinfo {title} {{Non-Abelian chiral instabilities at
  high temperature on the lattice}},\ }\href
  {https://doi.org/10.1007/JHEP03(2016)210} {\bibfield  {journal} {\bibinfo
  {journal} {JHEP}\ }\textbf {\bibinfo {volume} {03}},\ \bibinfo {pages}
  {210}},\ \Eprint {https://arxiv.org/abs/1512.02374} {arXiv:1512.02374
  [hep-ph]} \BibitemShut {NoStop}%
\bibitem [{\citenamefont {Hirono}\ \emph {et~al.}(2015)\citenamefont {Hirono},
  \citenamefont {Kharzeev},\ and\ \citenamefont {Yin}}]{Hirono:2015rla}%
  \BibitemOpen
  \bibfield  {author} {\bibinfo {author} {\bibfnamefont {Y.}~\bibnamefont
  {Hirono}}, \bibinfo {author} {\bibfnamefont {D.}~\bibnamefont {Kharzeev}},\
  and\ \bibinfo {author} {\bibfnamefont {Y.}~\bibnamefont {Yin}},\ }\bibfield
  {title} {\bibinfo {title} {{Self-similar inverse cascade of magnetic helicity
  driven by the chiral anomaly}},\ }\href
  {https://doi.org/10.1103/PhysRevD.92.125031} {\bibfield  {journal} {\bibinfo
  {journal} {Phys. Rev. D}\ }\textbf {\bibinfo {volume} {92}},\ \bibinfo
  {pages} {125031} (\bibinfo {year} {2015})},\ \Eprint
  {https://arxiv.org/abs/1509.07790} {arXiv:1509.07790 [hep-th]} \BibitemShut
  {NoStop}%
\bibitem [{\citenamefont {Xia}\ \emph {et~al.}(2016)\citenamefont {Xia},
  \citenamefont {Qin},\ and\ \citenamefont {Wang}}]{Xia:2016any}%
  \BibitemOpen
  \bibfield  {author} {\bibinfo {author} {\bibfnamefont {X.-l.}\ \bibnamefont
  {Xia}}, \bibinfo {author} {\bibfnamefont {H.}~\bibnamefont {Qin}},\ and\
  \bibinfo {author} {\bibfnamefont {Q.}~\bibnamefont {Wang}},\ }\bibfield
  {title} {\bibinfo {title} {{Approach to Chandrasekhar-Kendall-Woltjer State
  in a Chiral Plasma}},\ }\href {https://doi.org/10.1103/PhysRevD.94.054042}
  {\bibfield  {journal} {\bibinfo  {journal} {Phys. Rev. D}\ }\textbf {\bibinfo
  {volume} {94}},\ \bibinfo {pages} {054042} (\bibinfo {year} {2016})},\
  \Eprint {https://arxiv.org/abs/1607.01126} {arXiv:1607.01126 [nucl-th]}
  \BibitemShut {NoStop}%
\bibitem [{\citenamefont {Tuchin}(2018)}]{Tuchin:2017vwb}%
  \BibitemOpen
  \bibfield  {author} {\bibinfo {author} {\bibfnamefont {K.}~\bibnamefont
  {Tuchin}},\ }\bibfield  {title} {\bibinfo {title} {{Taming instability of
  magnetic field in chiral medium}},\ }\href
  {https://doi.org/10.1016/j.nuclphysa.2017.09.015} {\bibfield  {journal}
  {\bibinfo  {journal} {Nucl. Phys. A}\ }\textbf {\bibinfo {volume} {969}},\
  \bibinfo {pages} {1} (\bibinfo {year} {2018})},\ \Eprint
  {https://arxiv.org/abs/1702.07329} {arXiv:1702.07329 [nucl-th]} \BibitemShut
  {NoStop}%
\bibitem [{\citenamefont {Schlichting}\ and\ \citenamefont
  {Sharma}(2022)}]{Schlichting:2022fjc}%
  \BibitemOpen
  \bibfield  {author} {\bibinfo {author} {\bibfnamefont {S.}~\bibnamefont
  {Schlichting}}\ and\ \bibinfo {author} {\bibfnamefont {S.}~\bibnamefont
  {Sharma}},\ }\bibfield  {title} {\bibinfo {title} {{Chiral instabilities \&
  the fate of chirality imbalance in non-Abelian plasmas}},\ }\href@noop {} {\
  (\bibinfo {year} {2022})},\ \Eprint {https://arxiv.org/abs/2211.11365}
  {arXiv:2211.11365 [hep-ph]} \BibitemShut {NoStop}%
\bibitem [{\citenamefont {Qiu}\ \emph {et~al.}(2017)\citenamefont {Qiu},
  \citenamefont {Cao},\ and\ \citenamefont {Huang}}]{Qiu:2016hzd}%
  \BibitemOpen
  \bibfield  {author} {\bibinfo {author} {\bibfnamefont {Z.}~\bibnamefont
  {Qiu}}, \bibinfo {author} {\bibfnamefont {G.}~\bibnamefont {Cao}},\ and\
  \bibinfo {author} {\bibfnamefont {X.-G.}\ \bibnamefont {Huang}},\ }\bibfield
  {title} {\bibinfo {title} {{On electrodynamics of chiral matter}},\ }\href
  {https://doi.org/10.1103/PhysRevD.95.036002} {\bibfield  {journal} {\bibinfo
  {journal} {Phys. Rev. D}\ }\textbf {\bibinfo {volume} {95}},\ \bibinfo
  {pages} {036002} (\bibinfo {year} {2017})},\ \Eprint
  {https://arxiv.org/abs/1612.06364} {arXiv:1612.06364 [cond-mat.mes-hall]}
  \BibitemShut {NoStop}%
\bibitem [{\citenamefont {Galitski}\ \emph {et~al.}(2018)\citenamefont
  {Galitski}, \citenamefont {Kargarian},\ and\ \citenamefont
  {Syzranov}}]{Galitski:2018}%
  \BibitemOpen
  \bibfield  {author} {\bibinfo {author} {\bibfnamefont {V.}~\bibnamefont
  {Galitski}}, \bibinfo {author} {\bibfnamefont {M.}~\bibnamefont
  {Kargarian}},\ and\ \bibinfo {author} {\bibfnamefont {S.}~\bibnamefont
  {Syzranov}},\ }\bibfield  {title} {\bibinfo {title} {{Dynamo Effect and
  Turbulence in Hydrodynamic Weyl Metals}},\ }\href
  {https://doi.org/10.1103/PhysRevLett.121.176603} {\bibfield  {journal}
  {\bibinfo  {journal} {Phys. Rev. Lett.}\ }\textbf {\bibinfo {volume} {121}},\
  \bibinfo {pages} {176603} (\bibinfo {year} {2018})}\BibitemShut {NoStop}%
\bibitem [{\citenamefont {Amitani}\ and\ \citenamefont
  {Nishida}(2023)}]{Amitani:2022pmu}%
  \BibitemOpen
  \bibfield  {author} {\bibinfo {author} {\bibfnamefont {T.}~\bibnamefont
  {Amitani}}\ and\ \bibinfo {author} {\bibfnamefont {Y.}~\bibnamefont
  {Nishida}},\ }\bibfield  {title} {\bibinfo {title} {{Dynamical chiral
  magnetic current and instability in Weyl semimetals}},\ }\href
  {https://doi.org/10.1103/PhysRevB.107.014302} {\bibfield  {journal} {\bibinfo
   {journal} {Phys. Rev. B}\ }\textbf {\bibinfo {volume} {107}},\ \bibinfo
  {pages} {014302} (\bibinfo {year} {2023})},\ \Eprint
  {https://arxiv.org/abs/2207.14272} {arXiv:2207.14272 [cond-mat.mes-hall]}
  \BibitemShut {NoStop}%
\bibitem [{\citenamefont {Hattori}\ \emph {et~al.}(2019)\citenamefont
  {Hattori}, \citenamefont {Hirono}, \citenamefont {Yee},\ and\ \citenamefont
  {Yin}}]{Hattori:2017usa}%
  \BibitemOpen
  \bibfield  {author} {\bibinfo {author} {\bibfnamefont {K.}~\bibnamefont
  {Hattori}}, \bibinfo {author} {\bibfnamefont {Y.}~\bibnamefont {Hirono}},
  \bibinfo {author} {\bibfnamefont {H.-U.}\ \bibnamefont {Yee}},\ and\ \bibinfo
  {author} {\bibfnamefont {Y.}~\bibnamefont {Yin}},\ }\bibfield  {title}
  {\bibinfo {title} {{MagnetoHydrodynamics with chiral anomaly: phases of
  collective excitations and instabilities}},\ }\href
  {https://doi.org/10.1103/PhysRevD.100.065023} {\bibfield  {journal} {\bibinfo
   {journal} {Phys. Rev. D}\ }\textbf {\bibinfo {volume} {100}},\ \bibinfo
  {pages} {065023} (\bibinfo {year} {2019})},\ \Eprint
  {https://arxiv.org/abs/1711.08450} {arXiv:1711.08450 [hep-th]} \BibitemShut
  {NoStop}%
\bibitem [{\citenamefont {Yamamoto}\ and\ \citenamefont
  {Yang}(2021)}]{Yamamoto:2021gts}%
  \BibitemOpen
  \bibfield  {author} {\bibinfo {author} {\bibfnamefont {N.}~\bibnamefont
  {Yamamoto}}\ and\ \bibinfo {author} {\bibfnamefont {D.-L.}\ \bibnamefont
  {Yang}},\ }\bibfield  {title} {\bibinfo {title} {{Helical magnetic effect and
  the chiral anomaly}},\ }\href {https://doi.org/10.1103/PhysRevD.103.125003}
  {\bibfield  {journal} {\bibinfo  {journal} {Phys. Rev. D}\ }\textbf {\bibinfo
  {volume} {103}},\ \bibinfo {pages} {125003} (\bibinfo {year} {2021})},\
  \Eprint {https://arxiv.org/abs/2103.13208} {arXiv:2103.13208 [hep-th]}
  \BibitemShut {NoStop}%
\bibitem [{\citenamefont {Deng}\ and\ \citenamefont
  {Huang}(2016)}]{Deng:2016gyh}%
  \BibitemOpen
  \bibfield  {author} {\bibinfo {author} {\bibfnamefont {W.-T.}\ \bibnamefont
  {Deng}}\ and\ \bibinfo {author} {\bibfnamefont {X.-G.}\ \bibnamefont
  {Huang}},\ }\bibfield  {title} {\bibinfo {title} {{Vorticity in Heavy-Ion
  Collisions}},\ }\href {https://doi.org/10.1103/PhysRevC.93.064907} {\bibfield
   {journal} {\bibinfo  {journal} {Phys. Rev. C}\ }\textbf {\bibinfo {volume}
  {93}},\ \bibinfo {pages} {064907} (\bibinfo {year} {2016})},\ \Eprint
  {https://arxiv.org/abs/1603.06117} {arXiv:1603.06117 [nucl-th]} \BibitemShut
  {NoStop}%
\bibitem [{\citenamefont {Deng}\ \emph {et~al.}(2020)\citenamefont {Deng},
  \citenamefont {Huang}, \citenamefont {Ma},\ and\ \citenamefont
  {Zhang}}]{Deng:2020ygd}%
  \BibitemOpen
  \bibfield  {author} {\bibinfo {author} {\bibfnamefont {X.-G.}\ \bibnamefont
  {Deng}}, \bibinfo {author} {\bibfnamefont {X.-G.}\ \bibnamefont {Huang}},
  \bibinfo {author} {\bibfnamefont {Y.-G.}\ \bibnamefont {Ma}},\ and\ \bibinfo
  {author} {\bibfnamefont {S.}~\bibnamefont {Zhang}},\ }\bibfield  {title}
  {\bibinfo {title} {{Vorticity in low-energy heavy-ion collisions}},\ }\href
  {https://doi.org/10.1103/PhysRevC.101.064908} {\bibfield  {journal} {\bibinfo
   {journal} {Phys. Rev. C}\ }\textbf {\bibinfo {volume} {101}},\ \bibinfo
  {pages} {064908} (\bibinfo {year} {2020})},\ \Eprint
  {https://arxiv.org/abs/2001.01371} {arXiv:2001.01371 [nucl-th]} \BibitemShut
  {NoStop}%
\bibitem [{\citenamefont {Jiang}\ \emph {et~al.}(2016)\citenamefont {Jiang},
  \citenamefont {Lin},\ and\ \citenamefont {Liao}}]{Jiang:2016woz}%
  \BibitemOpen
  \bibfield  {author} {\bibinfo {author} {\bibfnamefont {Y.}~\bibnamefont
  {Jiang}}, \bibinfo {author} {\bibfnamefont {Z.-W.}\ \bibnamefont {Lin}},\
  and\ \bibinfo {author} {\bibfnamefont {J.}~\bibnamefont {Liao}},\ }\bibfield
  {title} {\bibinfo {title} {{Rotating quark-gluon plasma in relativistic heavy
  ion collisions}},\ }\href {https://doi.org/10.1103/PhysRevC.94.044910}
  {\bibfield  {journal} {\bibinfo  {journal} {Phys. Rev. C}\ }\textbf {\bibinfo
  {volume} {94}},\ \bibinfo {pages} {044910} (\bibinfo {year} {2016})},\
  \bibinfo {note} {[Erratum: Phys.Rev.C 95, 049904 (2017)]},\ \Eprint
  {https://arxiv.org/abs/1602.06580} {arXiv:1602.06580 [hep-ph]} \BibitemShut
  {NoStop}%
\bibitem [{\citenamefont {Son}\ and\ \citenamefont
  {Zhitnitsky}(2004)}]{Son:2004tq}%
  \BibitemOpen
  \bibfield  {author} {\bibinfo {author} {\bibfnamefont {D.~T.}\ \bibnamefont
  {Son}}\ and\ \bibinfo {author} {\bibfnamefont {A.~R.}\ \bibnamefont
  {Zhitnitsky}},\ }\bibfield  {title} {\bibinfo {title} {{Quantum anomalies in
  dense matter}},\ }\href {https://doi.org/10.1103/PhysRevD.70.074018}
  {\bibfield  {journal} {\bibinfo  {journal} {Phys. Rev. D}\ }\textbf {\bibinfo
  {volume} {70}},\ \bibinfo {pages} {074018} (\bibinfo {year} {2004})},\
  \Eprint {https://arxiv.org/abs/hep-ph/0405216} {arXiv:hep-ph/0405216}
  \BibitemShut {NoStop}%
\bibitem [{\citenamefont {Metlitski}\ and\ \citenamefont
  {Zhitnitsky}(2005)}]{Metlitski:2005pr}%
  \BibitemOpen
  \bibfield  {author} {\bibinfo {author} {\bibfnamefont {M.~A.}\ \bibnamefont
  {Metlitski}}\ and\ \bibinfo {author} {\bibfnamefont {A.~R.}\ \bibnamefont
  {Zhitnitsky}},\ }\bibfield  {title} {\bibinfo {title} {{Anomalous axion
  interactions and topological currents in dense matter}},\ }\href
  {https://doi.org/10.1103/PhysRevD.72.045011} {\bibfield  {journal} {\bibinfo
  {journal} {Phys. Rev. D}\ }\textbf {\bibinfo {volume} {72}},\ \bibinfo
  {pages} {045011} (\bibinfo {year} {2005})},\ \Eprint
  {https://arxiv.org/abs/hep-ph/0505072} {arXiv:hep-ph/0505072} \BibitemShut
  {NoStop}%
\bibitem [{\citenamefont {Elsasser}(1950)}]{Elsasser:1950zz}%
  \BibitemOpen
  \bibfield  {author} {\bibinfo {author} {\bibfnamefont {W.~M.}\ \bibnamefont
  {Elsasser}},\ }\bibfield  {title} {\bibinfo {title} {{The Hydromagnetic
  Equations}},\ }\href {https://doi.org/10.1103/PhysRev.79.183} {\bibfield
  {journal} {\bibinfo  {journal} {Phys. Rev.}\ }\textbf {\bibinfo {volume}
  {79}},\ \bibinfo {pages} {183} (\bibinfo {year} {1950})}\BibitemShut
  {NoStop}%
\bibitem [{\citenamefont {Woltjer}(1958)}]{Woltjer:1958}%
  \BibitemOpen
  \bibfield  {author} {\bibinfo {author} {\bibfnamefont {L.}~\bibnamefont
  {Woltjer}},\ }\bibfield  {title} {\bibinfo {title} {{A Theorem on Force-Free
  Magnetic Fields}},\ }\href {https://doi.org/10.1073/pnas.44.6.489} {\bibfield
   {journal} {\bibinfo  {journal} {PNAS}\ }\textbf {\bibinfo {volume} {44}},\
  \bibinfo {pages} {489} (\bibinfo {year} {1958})}\BibitemShut {NoStop}%
\bibitem [{\citenamefont {Matthaeus}\ \emph {et~al.}(2008)\citenamefont
  {Matthaeus}, \citenamefont {Pouquet}, \citenamefont {Mininni}, \citenamefont
  {Dmitruk},\ and\ \citenamefont {Breech}}]{Matthaeus:2007pd}%
  \BibitemOpen
  \bibfield  {author} {\bibinfo {author} {\bibfnamefont {W.~H.}\ \bibnamefont
  {Matthaeus}}, \bibinfo {author} {\bibfnamefont {A.}~\bibnamefont {Pouquet}},
  \bibinfo {author} {\bibfnamefont {P.~D.}\ \bibnamefont {Mininni}}, \bibinfo
  {author} {\bibfnamefont {P.}~\bibnamefont {Dmitruk}},\ and\ \bibinfo {author}
  {\bibfnamefont {B.}~\bibnamefont {Breech}},\ }\bibfield  {title} {\bibinfo
  {title} {{Rapid Alignment of Velocity and Magnetic Field in
  Magnetohydrodynamic Turbulence}},\ }\href
  {https://doi.org/10.1103/PhysRevLett.100.085003} {\bibfield  {journal}
  {\bibinfo  {journal} {Phys. Rev. Lett.}\ }\textbf {\bibinfo {volume} {100}},\
  \bibinfo {pages} {085003} (\bibinfo {year} {2008})},\ \Eprint
  {https://arxiv.org/abs/0708.0801} {arXiv:0708.0801 [astro-ph]} \BibitemShut
  {NoStop}%
\bibitem [{\citenamefont {Yokoi}(2013)}]{Yokoi:2013mja}%
  \BibitemOpen
  \bibfield  {author} {\bibinfo {author} {\bibfnamefont {N.}~\bibnamefont
  {Yokoi}},\ }\bibfield  {title} {\bibinfo {title} {{Cross helicity and related
  dynamo}},\ }\href {https://doi.org/10.1080/03091929.2012.754022} {\bibfield
  {journal} {\bibinfo  {journal} {Geophys. Astrophys. Fluid Dyn.}\ }\textbf
  {\bibinfo {volume} {107}},\ \bibinfo {pages} {114} (\bibinfo {year}
  {2013})},\ \Eprint {https://arxiv.org/abs/1306.6348} {arXiv:1306.6348
  [astro-ph.SR]} \BibitemShut {NoStop}%
\bibitem [{\citenamefont {Yoshizawa}(1998)}]{Yoshizawa:book}%
  \BibitemOpen
  \bibfield  {author} {\bibinfo {author} {\bibfnamefont {A.}~\bibnamefont
  {Yoshizawa}},\ }\href {https://doi.org/10.1017/9781316672853} {\emph
  {\bibinfo {title} {{Hydrodynamic and Magnetohydrodynamic Turbulent Flows:
  Modelling and Statistical Theory}}}}\ (\bibinfo  {publisher} {Springer},\
  \bibinfo {year} {1998})\BibitemShut {NoStop}%
\bibitem [{\citenamefont {Davidson}(2017)}]{Davidson:2017}%
  \BibitemOpen
  \bibfield  {author} {\bibinfo {author} {\bibfnamefont {P.}~\bibnamefont
  {Davidson}},\ }\href {https://doi.org/10.1007/978-94-017-1810-3} {\emph
  {\bibinfo {title} {{Introduction to Magnetohydrodynamics, 2nd Ed.}}}}\
  (\bibinfo  {publisher} {Cambridge University Press},\ \bibinfo {year}
  {2017})\BibitemShut {NoStop}%
\end{thebibliography}%

\end{document}